# Diffusion in the Presence of Correlated Returns in a Two-dimensional Energy Landscape and non-Monotonic Friction Dependence: Examination of Simulation Results by a Random Walk Model

**Subhajit Acharya and Biman Bagchi***

**Solid State and Structural Chemistry Unit, Indian Institute of Science, Bengaluru, Karnataka 560 012, India**

*Corresponding author's email: bbagchi@iisc.ac.in

## ABSTRACT

*Diffusion in a multidimensional energy surface with minima and barriers is a problem of importance in statistical mechanics and also has wide applications, such as protein folding. To understand it in such a system, we carry out theory and simulations of a tagged particle moving on a two-dimensional periodic potential energy surface, both in the presence and absence of noise. Langevin dynamics simulations at multiple temperatures are carried out to obtain the diffusion coefficient of a solute particle. Friction is varied from zero to large values. Diffusive motion emerges in the limit of long times, even in the absence of noise, although the trajectory is found to remain correlated over a long time. This correlation is manifested in correlated returns to the starting minima following a scattering by surrounding maxima. Noise destroys this correlation, induces chaos, and increases diffusion at small friction. Diffusion thus exhibits a non-monotonic friction dependence at the intermediate value of the damping, ultimately converging to our theoretically predicted value. The latter is obtained by using the well-established relation between diffusion and random walk. An excellent agreement is obtained between theory and simulations in the high friction limit, but not so in the intermediate regime. The rate of escape from one cell to another is obtained from the multidimensional rate theory of Langer. We find that enhanced dimensionality plays an important role. In order to quantify the effects of noise on the potential-imposed coherence on the trajectories, we calculate the Lyapunov exponent. At small friction values, the Lyapunov exponent mimics the friction dependence of the rate.*

# I. INTRODUCTION

The relationship between diffusion and friction was first derived by Einstein in his classic work on Brownian motion[1] in 1905, where the single-particle transport rate diffusion (D) was related to friction ($\zeta$) on the particle by $D = k_B T / \zeta$. This simple relation may define friction, particularly in liquids, aided further by the Langevin equation. In many studies, the above relation or its generalization is used to obtain friction, for example, to obtain the effects of friction and viscosity on the rate of a chemical reaction.[2,3]

We note that friction is a linear response function, and Einstein's relation is a statement of linear response theory.[2] These are expected to be valid across systems, although the nature of the friction can be complex in systems such as diffusion on an energy landscape with multiple maxima and minima. The motion of a particle in a potential energy landscape finds applications in many branches of physics, chemistry, and biology to discuss diverse problems.[4–6] In these examples, diffusion usually involves crossing barriers in the presence of noise.[7,8]

In recent years, there have been impressive developments in the numerical evaluation of the multidimensional free energy surface.[9–13] Many of these studies were motivated by the pioneering analytical studies of Onuchic and Wolynes, who modeled protein folding as a configuration space diffusion on a rugged two-dimensional free energy surface where the collective variables were the size of the protein and the contact order parameter.[4-6] Most of the recent numerical calculations, say as protein folding and protein association-dissociation reactions, have used metadynamics, umbrella sampling, and more sophisticated techniques and have concentrated mainly on two-dimensional free energy surfaces.[14,15] A substantial amount of discussion has focused on the criteria for selecting the appropriate choice of order parameters.[16,17] In our earlier work, we have discussed these aspects in the context of insulin dimer dissociation.[18,19] We observed that the choice of appropriate order parameters might depend on the stage of the complex reaction. While at significant separation, the center-to-center distance is the desired order parameter. More specific quantities, like contact between two helices, each from one monomer, provide a better description at close separation. A recent work discussed these aspects in the context of insulin dimer dissociation.[18,19]

*While substantial effort has been devoted to calculating the multidimensional free energy surface, less effort has been directed towards the calculation of the rate or diffusion.* The frictional effects also determine the rate along with the barrier height. As we have experienced in the study of one-dimensional activated barrier crossing dynamics where the frictional effects can reduce the rate by more than one order of magnitude, similar effects can be operative in multidimensional barrier crossing dynamics.[20]

In this work, we choose a simple Hamiltonian system where all the parameters required for the rate calculation are readily available, and accurate estimation of the rate constant from the simulation is also feasible. Sometimes, it becomes hard to determine the rate directly from simulation, such as in many complex biological systems, even after constructing the free energy surface as a function of the reaction coordinates. We need to rely on theoretical methods to obtain the rate in such cases.

A systematic study of the combined effects of noise and potential energy surface on diffusion was carried out by Festa and d'Agliano,[21] who considered only the overdamped limit for a periodic potential energy surface. In that particular case, they obtained an eigenvalue solution for the long-time diffusion coefficient, which is given by

$$D = \frac{1}{2}\left[\nabla_k \nabla_k \lambda(k,1)\right]_{k=0} \quad (1)$$

with eigenvalue taking the form $\lambda(k,\alpha) = \frac{k_B T}{\zeta}\Lambda(k,\alpha)$. Here $\Lambda$ represents the dimensionless non-negative parameter and depends on the characteristic nature of the potential energy surface. Here $\zeta$ denotes the friction coefficient. This theoretical study bears a resemblance to Langer's well-known analysis of rate across a barrier in a multidimensional potential energy surface.[22] In Langer's analysis, the rate is also given by an eigenvalue that defines the unstable mode.

In the present work, we study the diffusional dynamics of a Brownian particle in a periodic and continuous-potential-analog of regular Lorentz gas where the potential energy surface in two dimensions is given by[23]

$$V(x,y) = \varepsilon \left[ \cos\left(x + \frac{y}{\sqrt{3}}\right) + \cos\left(x - \frac{y}{\sqrt{3}}\right) + \cos\left(\frac{2y}{\sqrt{3}}\right) \right] \quad (2)$$

Here $\varepsilon$ is the energy scaling constant described later. The potential function is depicted in **Figure 1** below. It is evident that a study of diffusion in this periodic cosine potential can also be relevant to a variety of applications like super-ionic conductors, the motion of adsorbates on crystal surfaces, polymers diffusing at the interfaces, molecular graphene, etc.[24–26] Moreover, even such an apparently simple system can exhibit rich dynamical features, as we demonstrate here.

In this work, we ask the following questions.

(i) What are the nature of diffusion in the zero and very low friction limit when the migration of a tagged particle depends on the details of the potential energy surface? How to characterize the emergence of diffusion? For diffusion to emerge, the dynamical system must become chaotic.

(ii) Is it possible to measure chaos quantitatively for this system? Does the divergence rate between two trajectories in the phase space exhibit a similar dependence on friction like diffusion?

(iii) If diffusion is found to exist, can we describe it as a random walk model? It is common wisdom that diffusion should have an underlying random walk picture.

(iv)     How do we calculate the diffusion coefficient values analytically as we vary noise and temperature?

We obtain the answers to the above questions by using a mix of analytical and numerical techniques.

The rest of the article is organized as follows. In section II, we detail the studied system and simulation methodology. Section III.(a) presents the non-monotonic behavior of the noise-driven diffusion against friction. Section III.(b) measures the chaos by computing the Lyapunov exponent. Section III.(c) calculates the barrier crossing rate by employing different theoretical and numerical schemes. In section III.(d), we validate the random walk model in the presence of crossing and recrossing. A discussion on the exponential scaling relation between diffusion and entropy for a particle following Langevin dynamics is presented in section III.(e). Finally, in section IV, we summarize our work and draw some general conclusions

## II.   SYSTEMS AND SIMULATIONS

In our system, the Brownian particle moves on a two-dimensional periodic potential given by Eq.(2), and the equation of motion is governed by the ordinary Langevin equation [Eq.(3)]

$$\frac{dv}{dt} = -\zeta v + \frac{1}{m} F(x, y) + R(t) \qquad (3)$$

where $v = \frac{dr}{dt}$, $F(x,y)$ is the two-dimensional force, $m$ is the mass of the particle, $\zeta$ denotes friction coefficient, and the noise $R(t)$ satisfies the fluctuation-dissipation theorem.

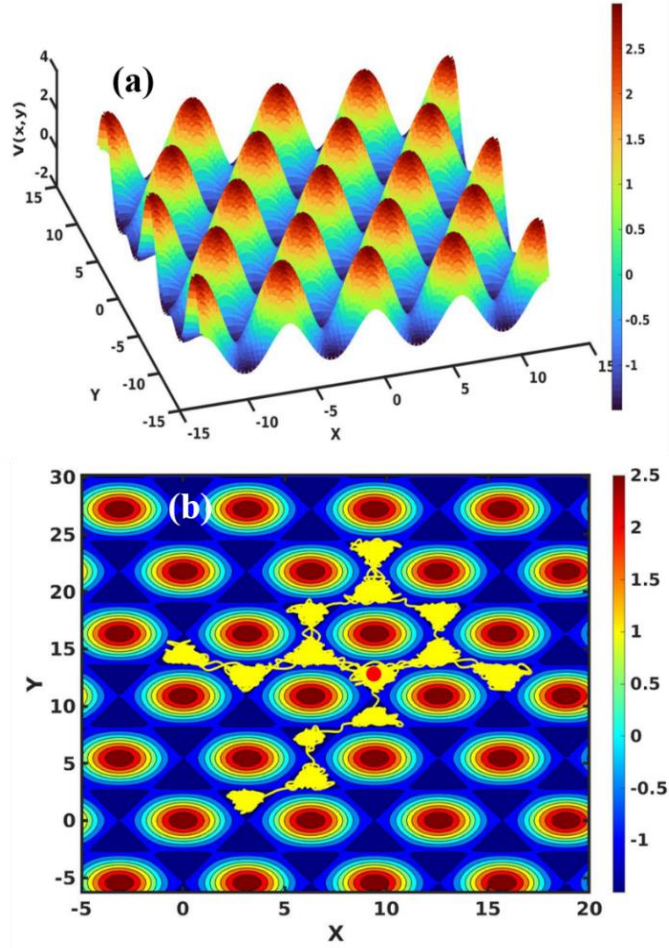

**Figure 1: (a) A schematic representation of the two-dimensional potential energy function defined in Eq.(2). The range and color codes are given on the right side of the plot. (b) The time evolution of a long trajectory is shown on this two-dimensional representation of potential energy function. Here the yellow line represents the trajectory of a Brownian particle starting at the minimum of a cell. The red circle represents the initial configuration of the particle in the phase space. Note that the trajectory makes multiple returns to the initial cell, indicating the presence of correlations, discussed further in the text.**

We project the potential energy function defined in Eq.(2) in the two-dimensional x-y plane and find that each cell contains a minimum at the center of the triangle, maxima at the three corners of the triangle, and saddle points at the mid-points of the edges, as shown in **Figure 1b**. Energies of the maxima, minimum, and saddle points of each cell are given by 3.0 $\varepsilon$, -1.5$\varepsilon$, and -1.0$\varepsilon$, respectively. We initiate the simulation by placing the Brownian particle near the minimum of the cell. We use Gear's fifth-order predictor-corrector algorithm to integrate the equation of motion of the particle with timestep 0.001 $\tau$ [27,28] where $\tau = \sqrt{\frac{m\sigma^2}{\varepsilon}}$ is the unit of time with *m* the mass, $\sigma$ the Lennard-Jones diameter, and $\varepsilon$ the Lennard-Jones well depth. In all our calculations, we use the reduced unit, such that the Lennard-Jones well depth $\varepsilon$ is the unit of energy and the Lennard-Jones diameter $\sigma$ is the unit of length. In the present study, we use the standard definition of reduced units. The calculations for 200 different initial configurations are performed to obtain a statistically significant and reproducible result at two reduced temperatures, T*=0.1 and T*=0.2.

## III.   RESULTS AND DISCUSSIONS

### a.   NON-MONOTONIC BEHAVIOR OF NOISE-DRIVEN DIFFUSION

The diffusional dynamics of a particle in this system are complex due to a refocusing of trajectories caused by a concave curvature in the potential energy surface, *as shown in **Figure 1b**. This is to be regarded as the opposite to dispersion caused by a convex surface*. This concavity causes an additional, long time (as opposed to short time, collisional) trapping of the trajectories, causing back-and-forth journeys between the same two cells. In a noiseless system where the existence of diffusion is solely the consequence of interactions with the potential energy surface, such refocusing lowers diffusion significantly.

We calculate the self-diffusion coefficient (D) of the Brownian particle using Einstein's relation between D and the mean square displacement (MSD). In two-dimension, self-diffusion coefficient is defined by

$$D = \lim_{t \to \infty} \frac{\langle (r(t) - r(0))^2 \rangle}{4t} \tag{4}$$

where *r*(t) is the position of the particle at time *t,* and angular brackets indicate the ensemble average. The obtained self-diffusion coefficient is plotted against noise in **Figure 2** at two reduced temperatures.

The results are expressed in dimensionless units where[29]

$$E^* = \frac{E}{\varepsilon}, \quad T^* = \frac{k_B T}{\varepsilon}, \quad \tau = \sqrt{\frac{m\sigma^2}{\varepsilon}}, \quad t^* = \frac{t}{\tau}, \quad D^* = \frac{D\tau}{\sigma^2}, \text{ and } \zeta^* = \zeta\tau$$

Here D* denotes the reduced self-diffusion coefficient, $\zeta$* denotes the reduced friction, E* is the reduced total energy, T* is the reduced temperature, and t* indicates the reduced time.

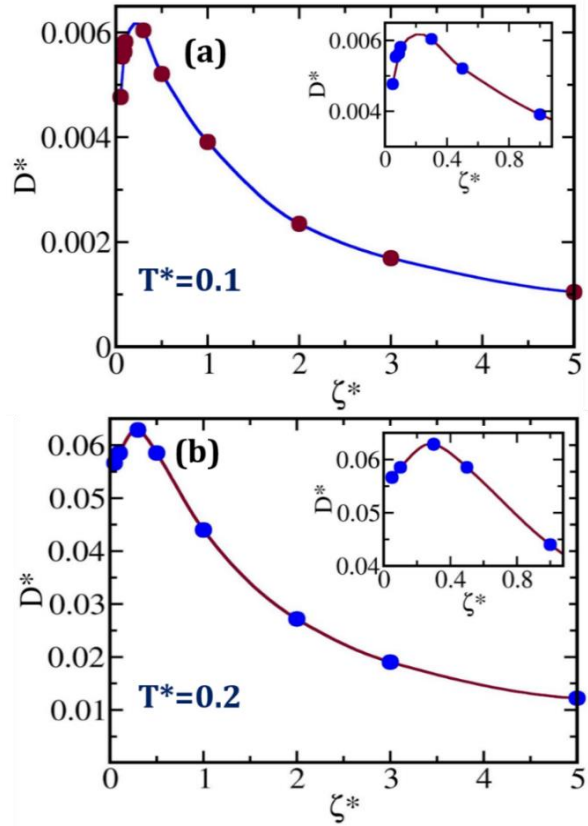

**Figure 2: Variation of self-diffusion coefficients against friction at two reduced temperatures (a) T*=0.1 and (b) T*=0.2. We estimate the self-diffusion coefficient of the Brownian particle by employing Eq.(4) and plot it against friction here. Diffusion exhibits a non-monotonic dependence on friction in both cases. Inset shows the variation of D* as a function of noise, specifically in the low friction region.**

We can obtain the self-diffusion coefficient also by integrating the un-normalized velocity autocorrelation. According to the Green-Kubo formalism, D is defined as,

$$D = \frac{1}{d}\int_0^\infty \langle \boldsymbol{v}(t).\boldsymbol{v}(0)\rangle dt \quad (5)$$

where *d* indicates the dimension of the system and *v*(t) is the velocity vector of the particle at time *t*. We plot the velocity auto-correlation function $C_v(t)$ against time at two reduced temperatures for different friction values in **Figure 3**. We observe that $C_v(t)$ for the noiseless system shows pronounced a long-time oscillatory behavior, which gradually decreases with noise. In the absence of noise, the particle gets trapped near the minimum of the initial triangular cell for a long time. Also, as mentioned above, it exhibits a coherence where the trajectory gets refocused after reaching the end of the adjacent cell and turns back to shuttle between the two cells. We will discuss these aspects later during trajectory analysis.

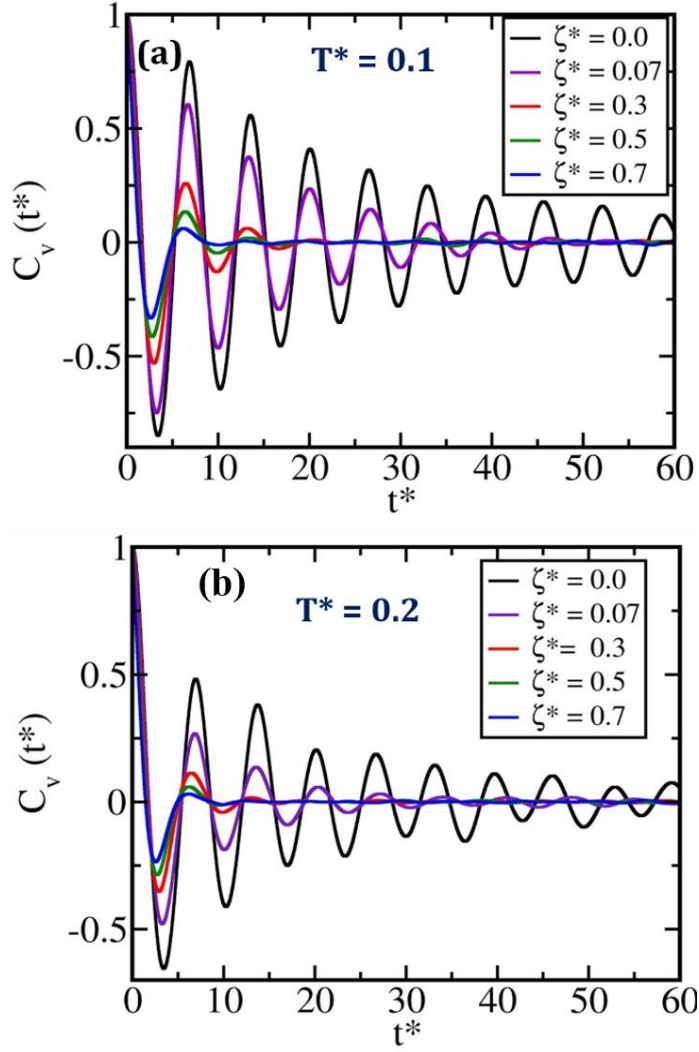

**Figure 3:** (a) The plot of the normalized velocity auto-correlation function $C_v(t)$ against time at a reduced temperature $T^*=0.1$. Here, we show the variation of $C_v(t)$ for several friction values (like, $\zeta^* = 0.0, \zeta^* = 0.07, \zeta^* = 0.3, \zeta^* = 0.5,$ and $\zeta^* = 0.7$). (b) The same is plotted at a reduced temperature $T^*=0.2$. In (a) and (b), we observe the oscillatory nature of $C_v(t)$ that is prominent in the absence of noise, gradually diminishes with the increase in strength of the noise.

**Figure 2** shows an interesting non-monotonic friction dependence of the diffusion at two different temperatures. This non-monotonic dependence of diffusion on friction has two non-trivial origins.

(i) Because of recrossing induced by the concave nature of the potential energy surface, diffusion is low in the absence of noise. Noise destroys this coherence and serves to increase diffusion. This is kind of novel and could be present in some cases.

(ii) A second reason (discussed more later) is the non-monotonic friction dependence of the barrier crossing rate. This is a two-dimensional version of Kramers' turnover, well-known in chemical kinetics.[30] These two are intimately connected, as discussed below.

b. **ONSET OF CHAOTIC MOTION AND CROSSOVER BEHAVIOR IN LYAPUNOV EXPONENT**

Trajectory analysis shows that our system undergoes both trapping near the minima and multiple crossing and recrossing through the saddle owing to the characteristic nature of the potential energy surface (as shown in **Figure 5**). Noise destroys the coherence by inducing chaos. Therefore, it is essential to study chaos quantitatively in this system.[31] In this regard, the Lyapunov exponent measures the average exponential rate of divergence or convergence of near orbits in the phase space.[32,33] Therefore, it is convenient to study the Lyapunov exponent ($L_n$) in order to understand the character of the motion. We employ "shadow trajectory" formalism to estimate $L_n$, as given by[33]

$$L_n = \frac{1}{n\Delta t} \sum_{i=1}^{n} \ln \frac{d_i}{d_{i-1}} \quad (6)$$

In order to implement this formalism, we initially choose two arbitrary close points separated by a distance ($d_0$) and monitor their mutual separation ($d_i$) with time. Here $d_i$ is the distance between two trajectories at $i^{th}$ timestep, and $d_{i-1}$ is the same at $(i-1)^{th}$ timestep. In Eq.(6), $n$ counts the step number and $\Delta t$ is the time interval between two steps. It is to be noted that, during the process, we need to renormalize the distance between the two selected trajectories periodically. *This procedure generates a time averaging over the two trajectories that start at close proximity in the phase space*. If we plot $L_n$ against $n$, it saturates after some steps, allowing $L_n$ to estimate unambiguously.

Figure 4 plots $L_n$ against friction at two reduced temperatures (i.e., T*=0.1 and T*=0.2). The calculated Lyapunov exponent exhibits an interesting dependence on noise. In Figure 4, we observe two significant trends. In the energy-controlled regime (i.e., at low friction), $L_n$ rises sharply with friction. However, in the intermediate to high friction region, the calculated $L_n$ exhibits a slow decay with the increase in friction. In the inset, we show the variation of $L_n$ with friction, emphasizing the overdamped limit. The inset shows that the divergence rate between two trajectories decreases with the increase in friction in the high friction limit.

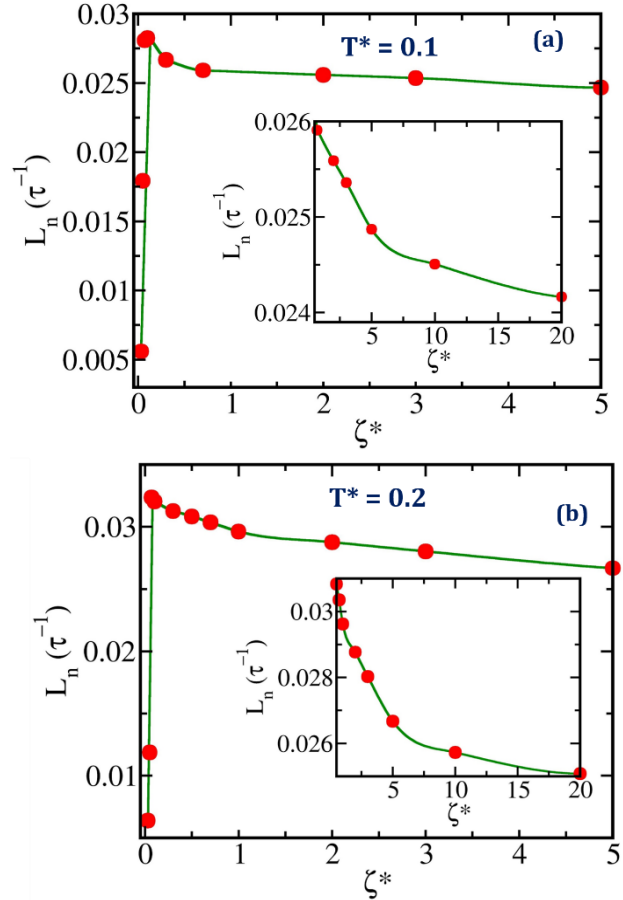

**Figure 4:** Plot of the calculated Lyapunov exponent ($L_n$) against friction at two reduced temperatures (a) T*=0.1 and (b) T*=0.2. Here we employ Eq.(6) to obtain the Lyapunov exponent for different friction values at two reduced temperatures. We observe that in the low damping regime, $L_n$ increases with friction sharply, whereas, in the overdamped regime, $L_n$ does not seem to increase rapidly due to the slowness of the medium. Inset shows the variation of $L_n$ against **friction, specifically in the high damping regime. In the overdamped limit, frictional effects favor the localization, which reduces the divergence rate of the two close trajectories in the phase space. This is why $L_n$ decreases with the increase in friction in this limit.**

In the energy-controlled regime, the motion of the Brownian particle is primarily dictated via collisions with the surrounding solvent molecules. Therefore, the motion becomes more chaotic with the slight increase in noise in this regime. However, beyond a particular value of friction, the system enters the diffusion-controlled regime where frictional effects slow down the motion of the particle. As discussed later, the analogy is very similar to Kramer's turnover problem. In this regime, the motion is also chaotic, and the frictional effects favor the localization of the trajectory in the phase space, which tends to reduce the Lyapunov exponent slightly. To make the discussion more concrete, we carry out trajectory analysis in **Figure 5** for some specific friction values. In the absence of noise and with an energy content below the saddle point energy, the particle remains trapped near the minimum of the initial cell, as shown in **Figure 5a**. In **Figure 5b**, we plot the trajectory of the Brownian particle at friction $\zeta^* = 0.3$. We choose this friction because, at this friction, diffusion becomes maximum, as shown in **Figure 2**.

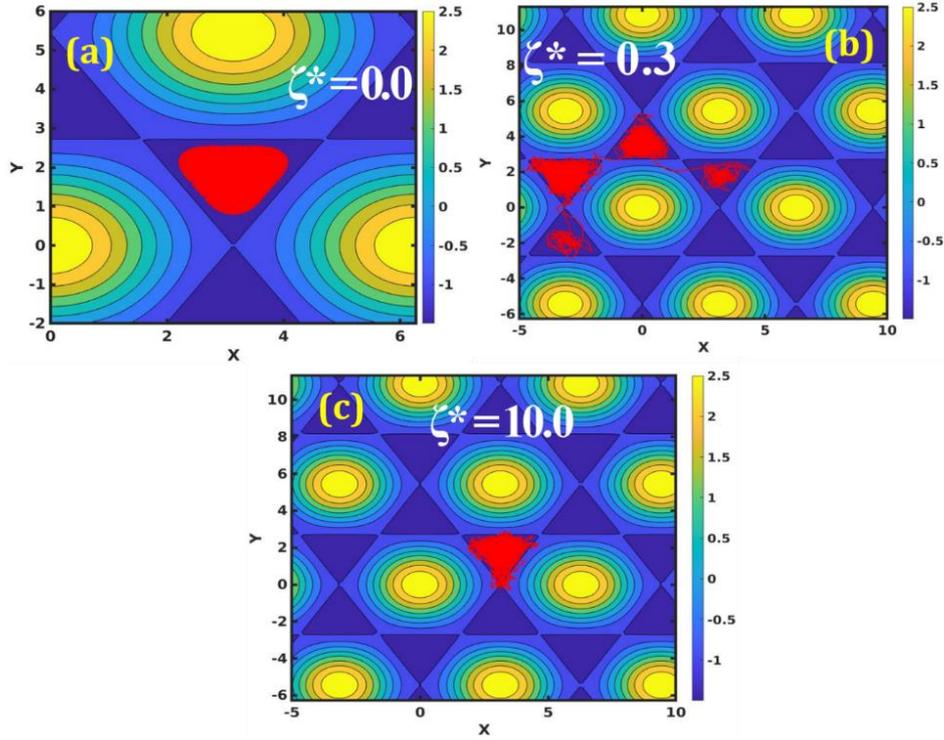

**Figure 5: The time evolution of the trajectory of the tagged particle on the two-dimensional potential energy surface shown by the red lines at a reduced temperature T\*=0.1. (a) In the absence of noise, the particle gets stuck inside the original cell, indicating that noise is mandatory for the motion of a particle. (b) We show the trajectory of the particle at friction $\zeta^* = 0.3$. We choose this particular friction because, at this friction, the diffusion becomes maximum, as discussed before. It clearly shows that the particle moves freely on the potential energy surface. (c) We demonstrate the trajectory of the particle in the comparably high friction limit (i.e., $\zeta^* = 10.0$ in this case).**

It is evident in **Figure 5b** that the particle explores the phase space at a faster rate than in the first case within the same time. In contrast, the particle in the overdamped limit (like at $\zeta^* = 10.0$) shows localization in the configurational phase space (as shown in **Figure 5c**).

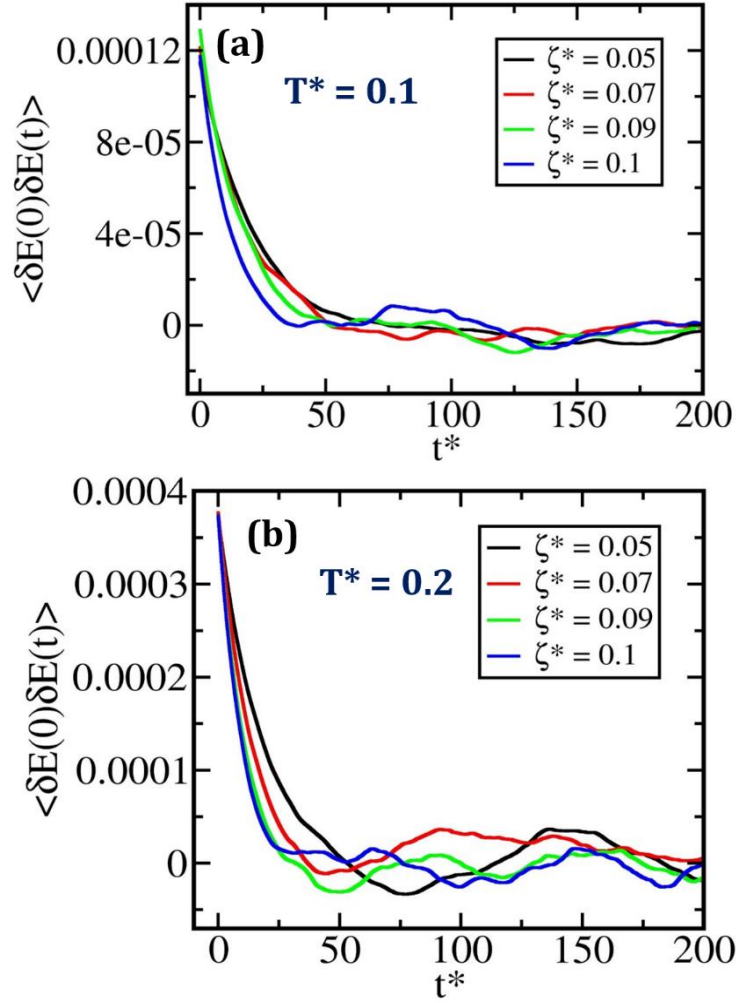

**Figure 6:** Variations of the un-normalized energy relaxation function against time at two reduced temperatures (a) T*=0.1 and (c) T*=0.2. It is observed that the relaxation of energy fluctuations $\langle \delta E(0)\delta E(t) \rangle$ becomes faster with the increase in friction in both cases. Exponential fitting parameters are provided in Table-1.

There is one more exciting aspect of the role of noise. As evident in the time trajectory of the total energy (i.e., **Figure 7**), the Brownian particle exhibits a non-zero escape rate even when the initial total energy of the particle is less than the saddle energy (i.e., $E_{saddle} = -1.0\varepsilon$ in this case). *This is a classic example of the role of noise.*

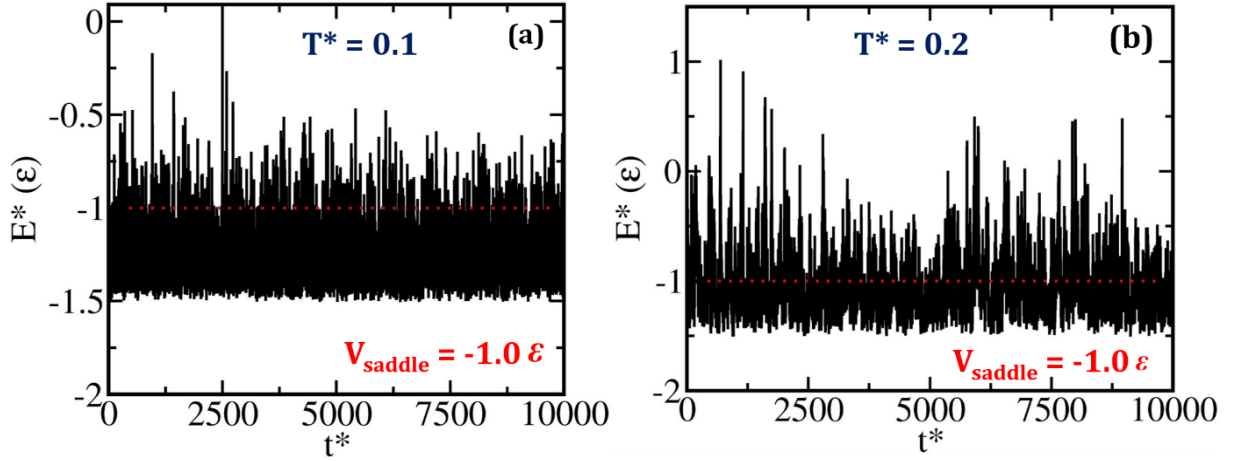

**Figure 7:** The plot of the total energy against time at temperatures (a) T*=0.1 and (b) T*=0.2. The red line indicates the potential energy near the saddle. When the total energy of the particle exceeds the saddle point energy, the particle can escape from the cell.

In **Figure 6**, we plot the fluctuation of energy relaxation $\langle \delta E(0)\delta E(t)\rangle$ against time where $\delta E(t)$ at time $t$ is given by $E(t)-\bar{E}$. Here $\bar{E}$ and $E(t)$ denote the average total energy and the total energy at time t for this system. We find that the correlation function for the fluctuations of energy relaxes faster with increasing noise strength noise induces dissipation and makes the system relax back to the equilibrium state at a faster rate.

**Table-1:** We use the exponential function $a_0 \exp\left(-\dfrac{t}{\tau}\right)$ to fit the normalized energy auto-correlation function at two reduced temperatures, T*=0.1 and T*=0.2. We report the fitting parameters here in reduced units.

| T* = 0.1 | | T* = 0.2 | |
|---|---|---|---|
| $\zeta^*$ | $\tau$ | $\zeta^*$ | $\tau$ |
| 0.05 | 18.57 | 0.05 | 15.67 |
| 0.07 | 16.12 | 0.07 | 11.49 |
| 0.09 | 14.54 | 0.09 | 8.80 |
| 0.1 | 10.88 | 0.1 | 8.70 |

We fit the normalized auto-correlation of energy fluctuation with the exponential function and report the fitting parameters in **Table-1** at two reduced temperatures (i.e., T*=0.1 and T*=0.2). According to the fluctuation-dissipation theorem, noise induces dissipation and makes the system relax back to the equilibrium state. We also study the power spectra of the total energy fluctuation, which provides a valuable window into the correlation present in the system in **Figure 8**. The power spectrum of the total energy fluctuation is given by

$$S_E(f) = \frac{1}{2\pi} \int_{-\infty}^{\infty} dt \exp(ift) \langle \delta E(0) \delta E(t) \rangle \tag{7}$$

The energy spectral density is found to be proportional $\frac{1}{f^\alpha}$, where $f$ is the frequency and $\alpha$ is the constant between 0.5 and 1.5. Such kind of behavior is known as 1/f noise, which indicates the possible intermittent transitions among multiple energy states of the system. [34–37]

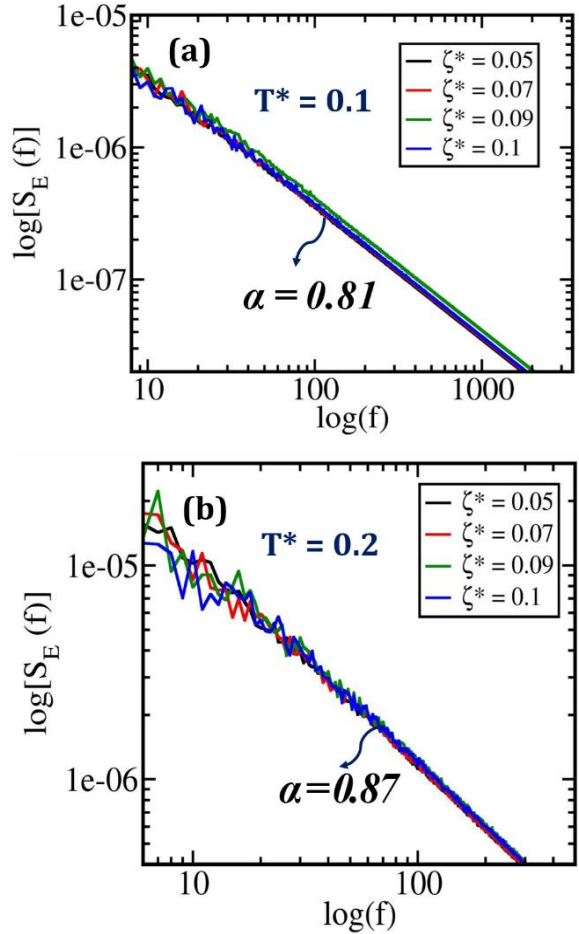

**Figure 8:** $1/f$ **noise behaviour of the spectral density is studied at two reduced temperatures (a) T\*=0.1 and (b) T\*=0.2. In (a) and (b), we plot the logarithm of the power spectrum of total energy fluctuations against the logarithm of the frequency.**

We plot such spectrum for the total energy of the particle in **Figures 8 (a)** and **(b)** at two reduced temperatures, T\*=0.1 and T\*=0.2. The origin of the slope in each case is mainly due to the effect of the solvent. As evident in **Figures 8 (a)** and **(b)**, frictional effects arising from solvent play an essential role in controlling both the energy spectrum and the dynamics of the particle.

**c.     CALCULATION OF THE ESCAPE RATE: COMPARISONS**

It is essential to estimate the escape rate of the tagged particle exhibiting the correlated random walk on a potential energy surface. Due to the simplicity of the chosen

Hamiltonian system, it is possible to compute the escape rate accurately by invoking both theoretical and simulation formalisms. In this section, we primarily study the escape rate variation obtained via several formalisms with friction.[38] In order to calculate the rate of a barrier crossing process, we require several parameters like well frequencies, barrier height, barrier frequency, etc.[22,30,39–42] Let us assume these in our isotropic two-dimensional system, X is the reaction coordinate, and Y is the nonreactive coordinate. We take the second derivative of the potential given by Eq.(2) with respect to the reactive and the nonreactive coordinates at the minimum of the initial cell and the saddle to extract well frequency, barrier frequency, etc. In **Table-2**, we report all the parameters required for the theoretical estimate of the escape rate in the reduced unit.

**Table-2: We report the parameters required to calculate the rate along the reactive and the nonreactive coordinates. Here, all the parameters are provided in the reduced unit.**

| | |
|---|---|
| Well frequency along X, $\omega_X^w$ | $1.0\,\tau^{-1}$ |
| Well frequency along Y, $\omega_Y^w$ | $1.0\,\tau^{-1}$ |
| Barrier frequency along X, $\omega_X^b$ | $0.82\,\tau^{-1}$ |
| Barrier frequency along Y, $\omega_Y^b$ | $1.41\,\tau^{-1}$ |
| Barrier height between the saddle and the minimum | $0.5\,\varepsilon$ |

There are several ways one can estimate the multidimensional rate theoretically.[42–45] In our study, we employ the multidimensional rate theory given by Langer to obtain the escape rate, which reads as [22,44]

$$k_L = \frac{\lambda_+}{2\pi}\left[\frac{\det E^w}{|\det E^b|}\right]^{1/2} \exp\left(-\frac{E^\dagger}{k_B T}\right) \tag{8}$$

where $\lambda_+$ is the only positive root of the equation $\det(\lambda I + E^b D) = 0$. Here D denotes the diffusion matrix, and $E^w$ and $E^b$ are the symmetric matrices containing the second partial derivatives of the free energy at the native well and saddle point, respectively. In Eq.(8) $E^\dagger$ denotes the activation barrier. We estimate the rate by employing Eq.(8) and plot it against friction in **Figure 9,** as shown by the green line at two different reduced temperatures, T*=0.1 and T*=0.2. In the limit of zero friction, Eq.(8) reduces to the transition state theory rate (given by Eq.(11)) $k_{TST}$, as shown by the black dotted line in **Figure 9**. Since transition state theory formalism neglects the effects of friction, $k_{TST}$ remains constant throughout the friction regime, as shown in **Figure 9**.

Now, we turn to compute the escape rate directly from the simulation. There are several ways one can estimate the rate directly from simulations.

(i) We can fit the population decay profile to a first-order exponential decay function. We shall refer to the rate by this procedure as the phenomenological rate.

(ii) One of the important methods to get the escape rate is the *Mean First Passage Time (MFPT) formalism.* In the mean first passage time (or first-order decay) formalism, we neglect the effects on the escape rate of the long-time correlated returns from the adjacent cells.[46] We, of course, include the immediate Kramers type recrossing by placing the barrier away from the saddle.

(iii) In order to take into account the effect long range correlations, one can employ the method first introduced by Yamamoto and popularized by Chandler, also used by Berne et al., etc.[47]

We know that recrossing through the saddle reduces the escape rate. However, here the recrossing has a different origin in our context.

According to the Yamamoto-Chandler formalism, we introduce a correlation function to take into account the effects of recrossing, which is defined as[47–49]

$$C(t) = \frac{\langle h(0)h(t) \rangle}{\langle h(0) \rangle} \tag{9}$$

Here $h(t)$ is a Heaviside function, which is one when the particle is inside the original cell and becomes zero whenever the particle is outside the original cell. It is noted that $h(0)$ always assumes the value of unity by this definition. For the rare transition between the two cells, $C(t)$ approaches as $C(t) \approx \langle h \rangle \left(1 - e^{-\frac{t}{\tau_{rxn}}}\right)$ where $\tau_{rxn} = (k_1 + k_{-1})^{-1}$. Here $k_1$ and $k_{-1}$ denote the forward rate constant and the backward rates, respectively. If the reaction time $\tau_{rxn}$ is much larger than the molecular time $\tau_{mol}$, there exists a time, i.e., $\tau_{mol} \prec t \ll \tau_{rxn}$ in which $C(t)$ grows linearly like $C(t) \sim k_1 t$. Consequently, reactive flux $k(t) = \frac{dC(t)}{dt}$ exhibits a plateau region whose value determines the forward rate constant $k_1$. According to this definition, we calculate the forward rate constant and plot against friction at two different reduced temperatures, as shown by the blue line in **Figure 9**. We observe that simulation results show a non-monotonic dependence of rate on friction, unlike theoretically predicted results.

    In the limit of zero friction, the transition state theory or Kramer's theory fails for the following reasons.[30,39,41] When the friction is very low, the system fails to acquire the energy required to cross the barrier as the sole energy source via colliding with the surrounding solvent molecules decreases significantly in this limit. Therefore, unlike the high friction limit, this limit has a proportional relation between friction and rate. Several attempts have been made to modify Kramer's theory to overcome the turnover issue in the low friction limit.[40,50] In this regard, Skinner and Wolynes derived an expansion of rate in the powers of friction coefficient, given by[51]

$$k = \frac{2\pi\zeta}{\omega_X^b} \left[ 1 + \frac{\frac{2\pi\zeta}{\omega_X^b}}{2} + \frac{\left(\frac{2\pi\zeta}{\omega_X^b}\right)^2}{2\pi} \right]^{-1} k_{TST} \qquad (10)$$

Here $k_{TST}$ denotes the two-dimensional transition state theory rate constant and is given by[52]

$$k_{TST} = \left(\frac{\omega_Y^w}{\omega_Y^b}\right) \frac{\omega_X^w}{2\pi} \exp\left(-\frac{E^\dagger}{k_B T}\right) \qquad (11)$$

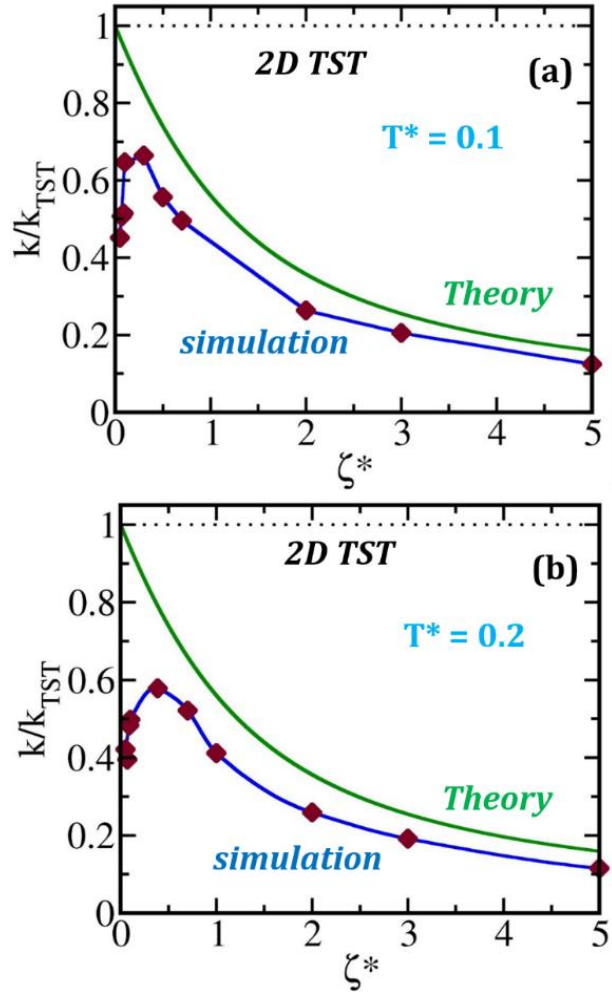

**Figure 9: Variation of the escape rate ($k/k_{TST}$) against friction at two reduced temperatures (a) T\*=0.1 and (b) T\*=0.2. Here $k_{TST}$ denotes the two-dimensional transition state theory rate given by Eq.(11) and is shown by the black dotted line. We use Langer's multidimensional rate theory to obtain the escape rate theoretically, as shown by the green line, to consider the effect of friction. The blue line denotes the rate obtained directly from simulation using population decay**

**formalism introduced by Yamamoto and Chandler. We have employed this formalism to consider the effect of long-time recrossing through the saddle.**

In Eq.(11), $\omega_X^w$ is the well frequency along the reactive coordinate X, $\omega_Y^w$ is the frequency near the well along the orthogonal nonreactive coordinate Y, and $\omega_Y^b$ denotes the harmonic frequency associated with the barrier along Y. In the overdamped limit, the multidimensional rate expression given by Eq.(8) reduces to the following equation

$$k^{SL} = \left(\frac{\omega_X^b}{\zeta}\right) k_{TST} \qquad (12)$$

Eq.(12) is popularly known as the Smoluchowski rate.[53] From **Figure 9**, it is clear that simulation results start converging to the theoretical predictions as we move to the overdamped regime. This is due to the reduction of the long-range correlated returns in the overdamped limit, discussed in the next section.

### d.    THE RANDOM WALK MODEL ESTIMATES AND COMPARISONS

This section compares the diffusion obtained via a regular random walk model with the same directly obtained from simulation. According to the regular random walk model, diffusion (D) is related to the rate constant (k) via the relation like[38]

$$D = \frac{1}{2d} ka^2,$$

Here, *d* denotes the dimension of the system, and *a* is the distance between the two adjacent minima.

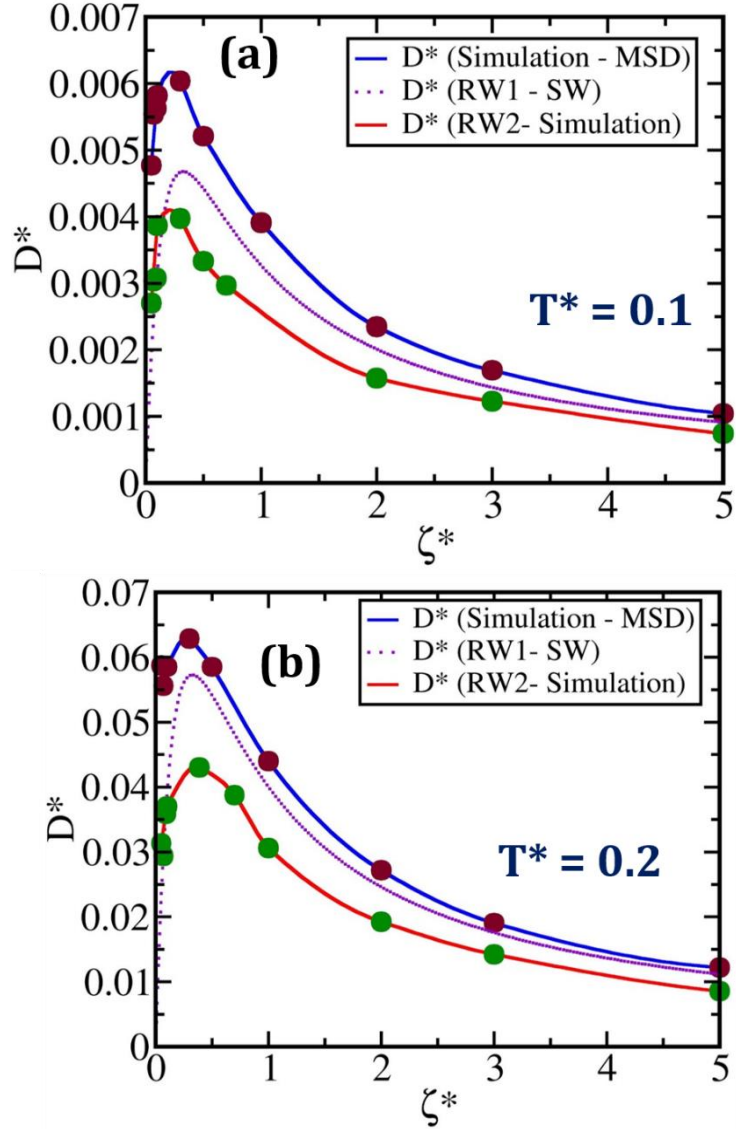

**Figure 10: Plot of self-diffusion coefficient against friction at two reduced temperatures (a) T\*=0.1 and (b) T\*=0.2. The blue line represents the variation of the diffusion against friction, obtained from the mean-square displacement (MSD) by the Langevin dynamics simulations. In theory, we employ the regular random walk model (RW), i.e., D=1/4 ka², to obtain the diffusion where the rate (k) is extracted from different theoretical formalisms. The purple line (Model RW1) shows the variation of diffusion against friction where *k* is predicted by Eq.(10) – the expression of Skinner & Wolynes. Similarly, we plot D obtained via the regular random walk model against friction, where k is calculated from simulation by employing the formalism introduced by Yamamoto and Chandler (model RW2, shown by the red line). From the plot, we observe that the random walk model predicts the diffusion well as we move to the overdamped limit. This is because, in the high friction limit, noise destroys the correlated motion of the particle. But the theories fail in the intermediate noise regime.**

On the other hand, we can directly calculate the escape rate ($k$) of the particle from simulation. We employ the formalism introduced by Yamamoto and Chandler to calculate the escape rate, considering the effects of correlated returns. We then extract the diffusion coefficient value by invoking the regular random walk model. The results are shown by the red line (model RW2) in **Figure 10**. We also obtain the diffusion using the regular random walk model where we use Eq.(11), which is the rate theory of Skinner-Wolynnes (SW).[51] The purple line represents the variation of this diffusion (model RW1) against friction. In **Figure 10**, the blue line shows the variation of diffusion against friction obtained directly from simulation via MSD.

The plots presented in **Figure 10** show *that the regular random walk model with the theoretically calculated rates predicts the diffusion satisfactorily only in the high friction regime*. This is due to the fact that as we move into the high friction regime, noise destroys the correlated motion of the particle. As discussed before, the dissipation of excess energy becomes faster in the overdamped limit. Therefore, the diffusion predicted by both random walk models (RW1 and RW2) shows acceptable convergence to the actual diffusion in this regime.

e.  **ENTROPY AND DIFFUSION**

There have been considerable discussions on the relationship between diffusion and entropy. Entropy is a measure of the configuration space available to the system, while diffusion is the rate of exploration of this configuration. In the liquid regime, this relation is straightforward, as discussed before. However, when diffusion occurs on a multidimensional energy landscape where the solute or the system has to cross barriers, the relation between diffusion and entropy becomes non-trivial. The regular random walk model allows us to use the relation between the rate constant ($k$) of crossing from one cell to the other to the diffusion coefficient (D) through the relation like, $D = \frac{1}{2d} k a^2$ where $a$ is lattice constant, the distance between two adjacent cells, and *d* represents the dimension of the system.[38] The above relation is valid only for the uncorrelated hopping between two traps when the configuration space is divisible into periodic cells.

On the other hand, according to the celebrated transition state theory of Wigner and Eyring, the rate constant is a property of phase space dynamics.[39,52,54] We use this connection to relate diffusion with entropy. In a recent article, we presented a microscopic derivation of the exponential relation between diffusion and entropy starting from the basic principles of Statistical Mechanics in both canonical and microcanonical ensembles.[55] In principle, we could attempt to derive an expression for the relationship between diffusion and entropy in a system where a tagged particle is coupled with a heat bath at temperature *T* and follows an ordinary Langevin equation defined by Eq.(3) with friction constant $\zeta$. This has not yet been accomplished. We hope to explore the relationship between diffusion and entropy by computing the configurational entropy for this system in the future.

## IV. CONCLUSION

The study of diffusion in a multidimensional surface has been a subject of immense interest in many branches of physics, chemistry, and biology to discuss diverse problems. In order to interrogate the role of noise in such diffusion processes, we study the diffusional dynamics of a Brownian particle in a periodic and continuous-potential-analog of regular Lorentz gas. The system of interest can also be relevant to a variety of applications like super-ionic conductors, the motion of adsorbates on crystal surfaces, polymers diffusing at the interfaces, molecular graphene, etc. We observe a non-monotonic dependence of diffusion as a function of friction in this two-dimensional periodic system. We explain the non-monotonic dependence of diffusion by invoking the random walk model with the rate given by an appropriate two-dimensional rate theory. The presence of coherence in the trajectory in the low friction limit gives rise to an exciting complexity that can be understood by using the Lyapunov exponent.

It is not clear *a priori* that a random walk model based analytical theory with an accurate rate calculation can reproduce the non-monotonic friction dependence of the diffusion constant. In fact, the Pade' approximant in the theory of Skinner-Wolynes is found to perform a reasonably satisfactory job in describing this non-monotonic dependence.

Although diffusion in the absence of noise is in itself a fascinating feature for this Hamiltonian system and has not really been studied in detail, our emphasis here is on the combined role of noise and energy landscape on diffusive motions. In the absence of noise, the particle gets trapped at the minimum of the initial cell and exhibits a pronounced oscillatory behavior in the plot of velocity auto-correlation against time. However, noise destroys the coherence by inducing chaos.

Trajectory analysis confirms that the particle suffers multiple long-range returns by crossing and recrossing the saddle in the low friction limit. Such long-range crossing and recrossing through saddle are different from those usually discussed in the context of barrier crossing dynamics, especially in relation to the theories of Kramers and Smoluchowski. The ones we focus on here return from other cells after rebounding from the maxima in other cells. These are not due to noise-generated solvent collisions.

In the presence of such correlated returns, the onset of chaotic motion is an interesting problem that needs to be studied. As we remarked earlier, the concave nature of the surface potential induces refocussing of the trajectories that leave one cell and enter the next adjoining cell. In order to continue the diffusive motion, the trajectory of the particle must become chaotic. In order to characterize the chaotic motion of the particle quantitatively, we calculate the Lyapunov exponent for this system and plot it against friction. We observe a nearly saturated region following a sharp rise in $L_n$ against friction in each case (i.e., T*=0.1 and T*=0.2) and address this issue in detail.

Although we have used a two-dimensional periodic landscape, the results are expected to remain valid in three and higher dimensions. Such generalization was observed earlier for diffusion on rugged energy landscapes.[56]

In a recent study, we attempted to analyze the rate of barrier crossing dynamics in a more complex two-dimensional reaction energy surface for insulin dimer dissociation.[18] In that case, a quantitatively accurate estimate of the rate required to validate the theoretical predictions was not available. This limitation led us to consider a simpler deterministic system where we can calculate not only the prediction of the rate by various theoretical approaches but also obtain a quantitatively accurate rate from simulations without making approximations.[57]

Diffusion here requires the escape of the tagged particle from one lattice cell to another one. We theoretically employ the multidimensional rate theory like the two-dimensional transition state theory and Langer's theory to compute the escape rate. In order to obtain the escape rate directly from simulation, we invoke the scheme developed by Yamamoto and Chandler, considering the effect of long-term recrossings. We find that Langer's theory can explain the high friction region semi-quantitatively but not the low friction domain. Thus, it fails to obtain the non-monotonic friction dependence.

We model the diffusion as a random walk in a two-dimensional lattice and invoke the relation between diffusion and rate (i.e., $D=1/4\ ka^2$) to calculate the diffusion with k obtained from the simulation. We compare that diffusion with the same directly obtained via MSD to validate the random walk model. With the increasing strength of the noise, the random walk model starts to predict a more accurate value for the diffusion since noise destroys the existing long-term correlated motions. Our study clearly shows the long-term correlated returns induced by the concave region of the surface play a significant role in lowering diffusion. In our study, we also demonstrate the origin of the agreement between simulation and theoretically predicted results in the overdamped limit.

**ACKNOWLEDGEMENTS**

BB thanks SERB (DST), India, for a National Science Chair (NSC) Professorship and SERB (DST), India, for partial funding of this work. SA thanks IISc for the research fellowship.

**References**

(1) Einstein, A. Investigations on the Theory of the Brownian Movement. *Ann. Phys.* **1905**, *17*, 549.

(2) Zwanzig, R. *Nonequilibrium Statistical Mechanics*; Oxford University Press, 2001.

(3) Bagchi, B. *Molecular Relaxation in Liquids*; Oxford Publications, 2012.

(4) Onuchic, J. N.; Luthey-Schulten, Z.; Wolynes, P. G. Theory of Protein Folding: The Energy Landscape Perspective. *Annu. Rev. Phys. Chem.* **1997**, *48* (1), 545–600. https://doi.org/10.1146/annurev.physchem.48.1.545.

(5) Bryngelson, J. D.; Wolynes, P. G. Intermediates and Barrier Crossing in a Random Energy Model (with Applications to Protein Folding). *J. Phys. Chem.* **1989**, *93* (19), 6902–6915. https://doi.org/10.1021/j100356a007.

(6) Bryngelson, J. D.; Wolynes, P. G. Spin Glasses and the Statistical Mechanics of Protein Folding. *Proc. Natl. Acad. Sci. U. S. A.* **1987**, *84* (21), 7524–7528. https://doi.org/10.1073/pnas.84.21.7524.

(7) Pierro, M. Di; Potoyan, D. A.; Wolynes, P. G.; Onuchic, J. N. Anomalous Diffusion, Spatial Coherence, and Viscoelasticity from the Energy Landscape of Human Chromosomes. *Proc. Natl. Acad. Sci. U. S. A.* **2018**, *115* (30), 7753–7758. https://doi.org/10.1073/pnas.1806297115.

(8) Socci, N. D.; Onuchic, J. N.; Wolynes, P. G. Diffusive Dynamics of the Reaction Coordinate for Protein Folding Funnels. *J. Chem. Phys.* **1996**, *104* (15), 5860–5868. https://doi.org/10.1063/1.471317.

(9) Bhimalapuram, P.; Chakrabarty, S.; Bagchi, B. Elucidating the Mechanism of Nucleation near the Gas-Liquid Spinodal. *Phys. Rev. Lett.* **2007**, *98* (20), 8–11. https://doi.org/10.1103/PhysRevLett.98.206104.

(10) Ghosh, R.; Roy, S.; Bagchi, B. Multidimensional Free Energy Surface of Unfolding of HP-36: Microscopic Origin of Ruggedness. *J. Chem. Phys.* **2014**, *141* (13). https://doi.org/10.1063/1.4896762.

(11) Hershkovitz, E.; Talkner, P.; Pollak, E.; Georgievskii, Y. Multiple Hops in Multidimensional Activated Surface Diffusion. *Surf. Sci.* **1999**, *421*, 73–88.

(12) Acharya, S.; Bagchi, B. Non-Markovian Rate Theory on a Multidimensional Reaction Surface: Complex Interplay between Enhanced Configuration Space and Memory. *J. Chem. Phys.* **2022**, *156* (13), 134101. https://doi.org/10.1063/5.0084146.

(13) Ma, A.; Nag, A.; Dinner, A. R. Dynamic Coupling between Coordinates in a Model for Biomolecular Isomerization. *J. Chem. Phys.* **2006**, *124* (14). https://doi.org/10.1063/1.2183768.

(14) Ayaz, C.; Tepper, L.; Brünig, F. N.; Kappler, J.; Daldrop, J. O.; Netz, R. R. Non-Markovian Modeling of Protein Folding. *Proc. Natl. Acad. Sci. U. S. A.* **2021**, *118* (31), 20–25. https://doi.org/10.1073/pnas.2023856118.

(15) Dama, J. F.; Rotskoff, G.; Parrinello, M.; Voth, G. A. Transition-Tempered Metadynamics: Robust, Convergent Metadynamics via on-the-Fly Transition Barrier Estimation. *J. Chem. Theory Comput.* **2014**, *10* (9), 3626–3633. https://doi.org/10.1021/ct500441q.

(16) Paul, S.; Taraphder, S. Determination of the Reaction Coordinate for a Key

Conformational Fluctuation in Human Carbonic Anhydrase II. *J. Phys. Chem. B* **2015**, *119* (34), 11403–11415. https://doi.org/10.1021/acs.jpcb.5b03655.

(17) Paul, S.; Paul, T. K.; Taraphder, S. Reaction Coordinate, Free Energy, and Rate of Intramolecular Proton Transfer in Human Carbonic Anhydrase II. *J. Phys. Chem. B* **2018**, *122* (11), 2851–2866. https://doi.org/10.1021/acs.jpcb.7b10713.

(18) Acharya, S.; Mondal, S.; Mukherjee, S.; Bagchi, B. Rate of Insulin Dimer Dissociation: Interplay between Memory Effects and Higher Dimensionality. *J. Phys. Chem. B* **2021**, *125* (34), 9678–9691. https://doi.org/10.1021/acs.jpcb.1c03779.

(19) Mukherjee, S.; Acharya, S.; Mondal, S.; Banerjee, P.; Bagchi, B. Structural Stability of Insulin Oligomers and Protein Association-Dissociation Processes: Free Energy Landscape and Universal Role of Water. *J. Phys. Chem. B* **2021**, *125* (43), 11793–11811. https://doi.org/10.1021/acs.jpcb.1c05811.

(20) Bagchi, B. Fractional Viscosity Dependence of Relaxation Rates and Non-Steady-State Dynamics in Barrierless Reactions in Solution. *Chem. Phys. Lett.* **1987**, *138* (4), 315–320. https://doi.org/10.1016/0009-2614(87)80390-1.

(21) Festa, R.; d'Agliano, E. G. Diffusion Coefficient for a Brownian Particle in a Periodic Field of Force. *Physica A* **1978**, *90A*, 229–244.

(22) Langer, J. S. Statistical Theory of the Decay of Metastable States. *Ann. Phys. (N. Y).* **1969**, *54*, 258–275.

(23) Bagchi, B.; Zwanzig, R.; Marchetti, M. C. Diffusion in a Two-Dimensional Periodic Potential. *Phys. Rev. A.* **1985**, *31* (2), 892–896.

(24) Ignatyuk, V. V. A Temperature Behavior of the Frustrated Translational Mode of Adsorbate and the Nature of the " Adsorbate – Substrate ." *J. Chem.Phys.* **2012**, *136* (2012), 184104.

(25) Challis, K. J.; Jack, M. W. Tight-Binding Approach to Overdamped Brownian Motion

on a Multidimensional Tilted Periodic Potential. *Phys. Rev. E.* **2013**, *87*, 052102. https://doi.org/10.1103/PhysRevE.87.052102.

(26) Gibertini, M.; Singha, A.; Pellegrini, V.; Polini, M.; Vignale, G.; Pinczuk, A.; Pfeiffer, L. N.; West, K. W. Engineering Artificial Graphene in a Two-Dimensional Electron Gas. *Phys. Rev. E.* **2009**, *79*, 241406. https://doi.org/10.1103/PhysRevB.79.241406.

(27) Ermak, D. L.; Buckholz, H. Numerical Integration of the Langevin Equation: Monte Carlo Simulation. *J. Comput. Phys.* **1980**, *35* (2), 169–182. https://doi.org/10.1016/0021-9991(80)90084-4.

(28) Bao, J. D.; Li, R. W.; Wu, W. Numerical Simulations of Generalized Langevin Equations with Deeply Asymptotic Parameters. *J. Comput. Phys.* **2004**, *197* (1), 241–252. https://doi.org/10.1016/j.jcp.2003.11.025.

(29) Allen, M. P.; Tildesley, D. J. *Computer Simulation of Liquids*; 1989; Vol. 45.

(30) Hanggi, P.; Talkner, P.; Borkovec, M. Reaction-Rate Theory : Fifty Years after Kramers. *Rev. Mod. Phys.* **1990**, *62* (2), 251–341.

(31) Gaspard, P.; Nicolis, G. Transport Properties, Lyapunov Exponents, and Entropy per Unit Time. *Phys. Rev. E.* **1990**, *65* (14), 1693–1696.

(32) Hofmann, T.; Merker, J. On Local Lyapunov Exponents of Chaotic Hamiltonian Systems. *CMST.* **2018**, *24* (2), 97–111. https://doi.org/10.12921/cmst.2017.0000053.

(33) Shevchenko, I. I.; Melnikov, A. V. Lyapunov Exponents in the Hénon – Heiles Problem. *JETP Lett.* **2003**, *77* (12), 642–646.

(34) Ohmine, I.; Saito, S. Water Dynamics: Fluctuation, Relaxation, and Chemical Reactions in Hydrogen Bond Network Rearrangement. *Acc. Chem. Res.* **1999**, *32* (9), 741–749. https://doi.org/10.1021/ar990107v.

(35) Ohmine, I.; Tanaka, H. Fluctuation, Relaxations, and Hydration in Liquid Water. Hydrogen-Bond Rearrangement Dynamics. *Chem. Rev.* **1993**, *93* (7), 2545–2566.

https://doi.org/10.1021/cr00023a011.

(36) Mukherjee, S.; Mondal, S.; Bagchi, B. Mechanism of Solvent Control of Protein Dynamics. *Phys. Rev. Lett.* **2019**, *122* (5), 58101. https://doi.org/10.1103/PhysRevLett.122.058101.

(37) Bagchi, B. *Water in Biological and Chemical Processes: From Structure and Dynamics to Function*; Cambridge University Press, 2013.

(38) Chandrasekhar, S. Reviews of Modern Physics. *Rev. Mod. Phys.* **1943**, *15*, 1–89.

(39) Eyring, H. The Activated Complex in Chemical Reactions. *J. Chem.Phys.* **1935**, *3*, 107–115.

(40) Grabert, H. Escape from a Metastable Well:The Kramers Turnover Problem. *Phys. Rev. Lett.* **1988**, *61* (3), 1683–1686.

(41) Kramers, H. A. Brownian Motion in a Field of Force and the Diffusion Model of Chemical Reactions. *Physica* **1940**, *7* (4), 284–304. https://doi.org/10.1016/S0031-8914(40)90098-2.

(42) Landauer, R.; Swanson, J. A. Frequency Factors in the Thermally Activated Process. *Phys. Rev.* **1961**, *121* (6), 1668–1674. https://doi.org/10.1103/PhysRev.121.1668.

(43) Van Der Zwan, G.; Hynes, J. T. Reactive Paths in the Diffusion Limit. *J. Chem. Phys.* **1982**, *77* (3), 1295–1301. https://doi.org/10.1063/1.443951.

(44) Langer, J. S. Theory of the Condensation Point. *Ann. Phys. (N. Y).* **1967**, *41*, 108–157.

(45) Truhlar, D. G.; Garrett, B. C. Multidimensional Transition State Theory and the Validity of Grote-Hynes Theory. *J. Phys. Chem. B* **2000**, *104* (5), 1069–1072. https://doi.org/10.1021/jp992430l.

(46) Müller, R.; Talkner, P.; Reimann, P. Rates and Mean First Passage Times. *Phys. A Stat. Mech. its Appl.* **1997**, *247* (1–4), 338–356. https://doi.org/10.1016/S0378-4371(97)00390-7.

(47) Montgomery, J. A.; Chandler, D.; Berne, B. J. Trajectory Analysis of a Kinetic Theory for Isomerization Dynamics in Condensed Phases. *J. Chem. Phys.* **1979**, *70* (9), 4056–4066. https://doi.org/10.1063/1.438028.

(48) Dellago, C.; Bolhuis, P. G.; Csajka, F. S.; Chandler, D. Transition Path Sampling and the Calculation of Rate Constants. *J. Chem. Phys.* **1998**, *108* (5), 1964–1977. https://doi.org/10.1063/1.475562.

(49) Dellago, C.; Bolhuis, P. G.; Chandler, D. On the Calculation of Reaction Rate Constants in the Transition Path Ensemble. *J. Chem. Phys.* **1999**, *110* (14), 6617–6625. https://doi.org/10.1063/1.478569.

(50) LEE, M.; HOLTOM, G. R.; Hochstrasser, R. M. Observation of The Kramers Turnover Region In The Isomerism of TRANS-STILBENE In Fluid Ethane. *Chem.Phys.Lett.* **1985**, *118* (4), 359–363.

(51) Skinner, J. L.; Woiynes, P. G. Relaxation Processes and Chemical Kinetics. *J. Chem. Phys.* **1978**, *69* (5), 2143–2150. https://doi.org/10.1063/1.436814.

(52) Pechukas, P. Transition State Theory. *Ann. Rev. Phys. Chem.* **1981**, *32*, 159–177.

(53) Smoluchowski, M. On the Kinetic Theory of the Brownian Molecular Motion and of Suspensions. *Ann. Phys.* **1906**, *21*.

(54) Laidler, K. J. *Chemical Kinetics*; Pearson Education, 1987.

(55) Acharya, S.; Bagchi, B. Study of Entropy-Diffusion Relation in Deterministic Hamiltonian Systems through Microscopic Analysis. *J. Chem. Phys.* **2020**, *153* (18). https://doi.org/10.1063/5.0022818.

(56) Seki, K.; Bagchi, K.; Bagchi, B. Anomalous Dimensionality Dependence of Diffusion in a Rugged Energy Landscape: How Pathological Is One Dimension? *J. Chem. Phys.* **2016**, *144* (19), 194106. https://doi.org/10.1063/1.4948936.

(57) Machta, J.; Zwanzig, R. Diffusion in a Periodic Lorentz Gas. *Phys. Rev. Lett.* **1983**, *50*

(25), 1959–1962.